\documentclass[a4paper,11pt]{article}
\pdfoutput=1 

\usepackage{jcappub} 

\usepackage[T1]{fontenc} 

\usepackage{graphicx}
\usepackage{amsmath}

\title{A ``nu'' look at gravitational waves: The black hole birth rate from neutrinos combined with the merger rate from LIGO}
\author{Jonathan H. Davis}
\author{and Malcolm Fairbairn}
\affiliation{Theoretical Particle Physics and Cosmology, Department of Physics, King's College London, London WC2R 2LS, United Kingdom}

\emailAdd{jonathan.davis@kcl.ac.uk}
\emailAdd{malcolm.fairbairn@kcl.ac.uk}

\abstract{We make projections for measuring the black hole birth rate from the diffuse supernova neutrino background (DSNB) by future neutrino experiments, and constrain the black hole merger fraction $\epsilon$, when combined
with information on the black hole merger rate from gravitational wave experiments such as LIGO. The DSNB originates from neutrinos emitted by all the supernovae in the Universe, and is expected to be made up of two components: neutrinos from neutron-star-forming supernovae,
and a sub-dominant component at higher energies from black-hole-forming ``unnovae''. We perform a Markov Chain Monte Carlo analysis of simulated
data of the DSNB in an experiment similar to Hyper-Kamiokande, focusing on this second component. Since all knowledge of the neutrino emission from unnovae comes from simulations of collapsing stars, we choose two sets of priors: one where the unnovae are well-understood and one where
their neutrino emission is poorly known. By combining the black hole birth rate from the DSNB with projected measurements of the black hole merger rate from LIGO, we show that the fraction of black holes which lead to binary mergers observed today $\epsilon$
could be constrained to be within the range $2 \cdot 10^{-4} \leq \epsilon \leq 3 \cdot 10^{-2}$ at $3 \sigma$ confidence,
after ten years of running an experiment like Hyper-Kamiokande.}

\begin{document}

\maketitle
\flushbottom

\section{Introduction}
For stars with masses $M$ above approximately $8 M_{\odot}$ the end of their fuel-burning phase results in a phenomenon known as core-collapse~\cite{Janka:2006fh}. At this point pressure from nuclear fusion can no longer counter the force of gravity and so the star's core contracts.  The end of this process is a violent supernova (SN), leading to a vast amount of energy released in the form of photons and neutrinos~\cite{Lang:2016zhv}. It is the neutrinos which carry away the majority of the released energy,
with over $10^{53}$~ergs emitted in neutrinos from a core-collapse supernova~\cite{Lunardini:2009ya,2010PhRvL.104y1101H}.

The flux and energy of this emission during this core-collapse should depend on the initial mass of the star. For stars with mass $8 M_{\odot} \lesssim M \lesssim 25 M_{\odot}$ the collapse will result in a neutron star (NS)
and a large flux of neutrino and photonic emission~\cite{Lunardini:2009ya}. However for more massive stars with $M \gtrsim 25 M_{\odot} \mbox{-} 40 M_{\odot}$ the core will collapse to a black hole (BH), with potentially different physics
as a result, leading to what is sometimes called an unnova~\cite{Yuksel:2012zy}. 
Simulations have shown that during such a collapse the flux of neutrinos is greater than for the NS-forming supernova events, and their average energy is larger~\cite{Lunardini:2009ya,PhysRevD.78.083014,Fischer:2008rh,Sumiyoshi:2008zw}.
There may also be considerably less photonic emission making such events difficult to observe with conventional telescopes, and indeed the only optical method of looking for unnovae may be to search for stars disappearing from the sky~\cite{Adams:2016ffj}.
Hence observations of the neutrino flux from one or more unnova events are likely to be the only way to obtain valuable information on the formation of black holes from core-collapse of stars, especially at larger redshifts up to $z \sim 1$ and
above, where optical disappearance searches are more difficult~\cite{Adams:2016hit}.

Neutrinos from supernovae are detectable in two broad categories: through the direct observation of a neutrino burst, potentially correlated with an optical event, and through the (as yet undetected but expected) diffuse supernova neutrino background (DSNB)~\cite{Beacom:2010kk,Lunardini:2010ab}. 
The detection of a neutrino burst within our own galaxy will lead to potentially $10^4$ neutrino detections within a short space of time in e.g. Super Kamiokande~\cite{Fukuda:2002uc,Ando:2005ka}, however we expect at most a few such galactic supernovae per
century~\cite{Ando:2005ka}. Indeed a burst of neutrinos from a supernova has been observed only once so far, from SN1987A~\cite{Hirata:1987hu}. The next generation of more massive detectors e.g. Hyper-Kamiokande~\cite{Abe:2011ts} will potentially be sensitive to neutrino bursts from supernovae up to a few Mpc away, which will occur more frequently, however the flux from such events on Earth
will be small and so the statistics will be limited~\cite{Ando:2005ka}.

By contrast the DSNB represents a continuous source of neutrinos from all of the supernovae which have occurred in the Universe, and does not require us to wait for a nearby supernova event to occur. If the simulations of BH-forming
collapses are correct, then this DSNB should be comprised broadly of two components: a larger flux from NS-forming core-collapse supernovae and a smaller component at higher-energies from BH-forming unnovae~\cite{Lunardini:2009ya,Yuksel:2012zy,Keehn:2010pn,Lien:2010yb,Nakazato:2015rya,Lunardini:2010ab}.
As pointed out in refs.~\cite{Lunardini:2009ya,Yuksel:2012zy,Keehn:2010pn,Lien:2010yb,Nakazato:2015rya,Lunardini:2010ab} by measuring the spectra and fluxes of both components we can obtian useful information on the birth rate of black holes as a function of redshift,
and so the DSNB is potentially a unique window into the physics of black hole production from stellar collapse. However this is made significantly more complicated by the large number of parameters which enter the calculation of the
DSNB, such as the redshift-dependent star formation rate, the flux and spectra of both NS-forming supernovae and BH-forming unnovae and the redshift dependence of the fraction of stars which collapse to NS or BH.
No comprehensive study looking at the degeneracies between all these parameters has yet been done, something which we address in this work.

Of all the black holes born long enough ago such that they have had enough time to lead to merger events today, only a fraction $\epsilon$ will exist in binary systems with the right properties to result in the production and observation of gravitational waves today, such as the event observed by the LIGO experiment towards the end of 2015~\cite{Abbott:2016blz,TheLIGOScientific:2016htt}.
 Indeed the LIGO experiment currently sets the merger rate of black hole binaries $\mathcal{R}_{\mathrm{BH-BH}}$ within the range $9 - 240 \, \mathrm{Gpc}^{-3} \mathrm{yr}^{-1}$ at $90\%$ confidence~\cite{TheLIGOScientific:2016pea}, and this should improve in precision significantly over the next decade~\cite{Kovetz:2016kpi}.
 This quantity $\epsilon$ is not well-known, though efforts have been made to infer its value using theoretical predictions for the black hole birth rate combined with LIGO data on the merger rate of black hole binaries~\cite{Elbert:2017sbr}. It should however be possible instead to use neutrinos from unnovae to constrain the black hole birth rate, and so place bounds on $\epsilon$.

In this work we discuss the potential of future observations using the Hyper-Kamiokande experiment to place constraints on the birth rate of black holes using the DSNB, and then show that by combining this with the merger rate of black holes from LIGO it is possible to place limits on the fraction of black holes which end up in binary mergers $\epsilon$.
In section~\ref{sec:dsnb_mcmc} we obtain robust projected constraints on the black hole birth rate from the DSNB for the upcoming Hyper-Kamiokande experiment by performing a Markov Chain Monte Carlo (MCMC) analysis, taking into
account all of the relevant nuisance parameters. 
Crucially we consider the fact that simulations of neutrino production during BH-forming collapse events may not accurately reflect the true physics, by using different prior distributions for the neutrino spectra from unnovae. 
We then combine this birth rate with the expectations and measurements of the BH-BH merger rate from LIGO in section~\ref{sec:ligo} to infer the fraction of black holes which lead to merger events $\epsilon$. We conclude in section~\ref{sec:conc}.

\section{MCMC study of the DSNB\label{sec:dsnb_mcmc}}
In this section, we seek to constrain the black hole birth rate $R_{\mathrm{BH}}(z)$ as a function of redshift $z$ by using projected measurements of the DSNB.
Previous studies have considered the effect of $R_{\mathrm{BH}}(z)$ on the DSNB~\cite{Lunardini:2009ya,Yuksel:2012zy,Keehn:2010pn,Lien:2010yb,Nakazato:2015rya,Lunardini:2010ab}, however in this work we seek instead to infer 
$R_{\mathrm{BH}}(z)$ from the DSNB, given the  full set of potentially degenerate parameters  e.g. the supernova and unnova spectrum. Once we have projected bounds on $R_{\mathrm{BH}}(z)$, we will be equipped to combine this information
with LIGO data on black hole mergers to infer the merger fraction.
Hence we perform a Markov Chain Monte Carlo analysis (MCMC) over all parameters on which the DSNB flux depends (see appendix~\ref{app:MCMC} for more information),
to understand to what extent $R_{\mathrm{BH}}(z)$ can be inferred from the DSNB given our potentially uncertain knowledge of the physics of supernovae and unnovae.
Importantly, since the only information we have for neutrino emission from unnovae is from simulations~\cite{PhysRevD.78.083014,Fischer:2008rh,Sumiyoshi:2008zw,Lunardini:2009ya}, we need to incorporate the possibility that unnovae do not exist,
or are identical to supernovae in their neutrino emission.

\subsection{Detecting the DSNB in a water Cherenkov experiment}

The DSNB is a flux of neutrinos from all of the supernovae (and potentially unnovae) which have occurred throughout the Universe. We focus on anti-neutrinos here as they are easier to detect.
The DSNB flux $\Phi(E)$ as a function of neutrino energy $E$ is calculated using the following integral over redshift $z$ with $15$
parameters which we list in table~\ref{table_params}. In each case a subscript ``NS'' refers to neutron star forming core-collapse events while subscript ``BH'' refers to black hole forming unnovae.
\begin{equation}
\begin{split}
\Phi(E) = \frac{c}{H_0} \int_0^{z_{\mathrm{max}}} {\frac{\mathrm{d} z}{\sqrt{\Omega_m (1+z)^3 + \Omega_{\Lambda}}}} \big[ R_{\mathrm{NS}}(z) F_{\mathrm{NS}}(E(1+z);\bar{E}_{e \mathrm{NS}},\bar{E}_{x \mathrm{NS}},L_{e \mathrm{NS}},L_{x \mathrm{NS}}) \\
+ R_{\mathrm{BH}}(z) F_{\mathrm{BH}}(E(1+z);\bar{E}_{e \mathrm{BH}},\bar{E}_{x \mathrm{BH}},L_{e \mathrm{BH}},L_{x \mathrm{BH}}) \big]
\label{eqn:dsnb_all}
\end{split}
\end{equation}
where $R_{\mathrm{NS}}(z)$ and $R_{\mathrm{BH}}(z)$ are respectively the redshift-dependent rates of NS and BH forming core-collapse events, $F_{\mathrm{NS}}$ and $F_{\mathrm{BH}}$ are the spectra of either supernovae or unnovae, $H_0$ is the Hubble constant, $c$ is the speed of light,
$\Omega_m \approx 0.3$ is the fractional matter density of the Universe and $\Omega_{\Lambda} \approx 0.7$ is the same for dark energy~\cite{Lunardini:2009ya,Yuksel:2012zy}. Here we take the maximum redshift to be $z_{\mathrm{max}} = 4$.

\begin{table}[t]
\begin{center}
\begin{tabular}{ c || c  }
\textbf{Parameter} & \textbf{Description}  \\ \hline 
$\bar{E}_{e \mathrm{NS}}$ and $\bar{E}_{e \mathrm{BH}}$& Average energy of $\bar{\nu}_e$\\
\hline
$\bar{E}_{x \mathrm{NS}}$ and $\bar{E}_{x \mathrm{BH}}$& Average energy of $\bar{\nu}_{\mu}$  and $\bar{\nu}_{\tau}$\\
\hline
${L}_{e \mathrm{NS}}$ and ${L}_{e \mathrm{BH}}$& Luminosity of $\bar{\nu}_e$\\
\hline
${L}_{x \mathrm{NS}}$ and ${L}_{x \mathrm{BH}}$& Luminosity of $\bar{\nu}_{\mu}$  and $\bar{\nu}_{\tau}$\\
\hline
$R_0$& Total rate of core-collapse events \\
&  at the present-day\\ \hline
$\gamma$ and $\beta$& Powers of redshift-dependent \\
 & total core-collapse rate \\ \hline
$f_0$, $f_1$ and $f_4$& Fraction of core-collapses which lead to BH  \\
&production at $z = 0$, $z = 1$ and $z = 4$ respectively \\ \hline
$\bar{p}$&Flavour oscillation parameter
\end{tabular}
\end{center}
\caption{List of parameters used in our analysis of the DSNB.}
\label{table_params}
\end{table}

Anti-neutrinos and neutrinos decouple from matter at a radial surface known as the neutrino-sphere~\cite{Sigl:1994da}, and their spectrum becomes an approximately thermal one with a fixed average energy $\bar{E}$. The value of $\bar{E}$ is different for $\nu_e$, $\bar{\nu}_e$
and the remaining heavy $\mu$ and $\tau$ flavours, which we denote as $x$, since each of these three types of neutrino interacts with matter with different strengths. Hence the $\bar{\nu}_e$ have a different spectrum from $\bar{\nu}_x$.

Since neutrinos oscillate between flavours after they are produced, a neutrino produced as a $\bar{\nu}_x$ may oscillate into a $\bar{\nu}_e$ or vice versa.
Due to the extremely small size of the neutrino wave-packet, most of this conversion will occur due to matter effects, and not as the neutrinos travel to Earth~\cite{Kersten:2015kio,PhysRevD.37.552}.
\footnote{This is a subtle point and follows from the fact that the neutrinos are produced from charged particles such as nuclei or electrons/positrons, which have an extremely small mean free path for scattering in the supernova.
Hence the average time over which neutrinos are emitted coherently is tiny, and so the wave-packet size can be as small as $10^{-11}$~cm~\cite{Kersten:2015kio,PhysRevD.37.552}. It follows therefore from the Pauli exclusion principle that the uncertainty on the neutrino
momentum and energy is large, and so we can not describe the different mass eigenstates of the neutrinos as propagating with different velocities. Hence there is practically no observable separation of mass eigenstates as the neutrinos travel to Earth (see also ref.~\cite{Wright:2017jwl}).} 
Indeed, the densities of matter in the remnant star during neutrino
production are so large that matter effects can dominate the flavour oscillations. In this work we follow ref.~\cite{Lunardini:2009ya} and parametrise the effect of flavour oscillations with the variable $\bar{p}$,
which can be anywhere between $0$ and $\cos^2 \theta_{12} = 0.68$. We use the same variable for both supernovae and unnovae, however in principle both such scenarios could have different prior distributions of $\bar{p}$, which may or 
may not be correlated. Our knowledge of $\bar{p}$ for supernovae and unnovae should improve greatly in the near future due to various upcoming measurements, such as a determination of the neutrino mass hierarchy, a better theoretical understanding of neutrino oscillations near the neutrinosphere and better measurements of the supernova neutrino spectrum~\cite{Mirizzi:2015eza}.
Hence, the spectrum of electron anti-neutrinos $\bar{\nu}_e$ which will be detected on Earth takes the form,
\begin{eqnarray}
F_{\mathrm{NS}} &=& \bar{p} \cdot J_{e \mathrm{NS}}(E(1+z);\bar{E}_{e \mathrm{NS}},L_{e \mathrm{NS}}) + (1 - \bar{p}) \cdot J_{x \mathrm{NS}}(E(1+z);\bar{E}_{x \mathrm{NS}},L_{x \mathrm{NS}}) \\
F_{\mathrm{BH}}&=& \bar{p} \cdot J_{e \mathrm{BH}}(E(1+z);\bar{E}_{e \mathrm{BH}},L_{e \mathrm{BH}}) + (1 - \bar{p}) \cdot J_{x \mathrm{BH}}(E(1+z);\bar{E}_{x \mathrm{BH}},L_{x \mathrm{BH}})
\end{eqnarray}
\begin{equation}
J(E,\bar{E},L) = \frac{L \cdot (1+\alpha)^{1+\alpha}}{\Gamma(1+\alpha) \bar{E}^2} \left( \frac{E}{\bar{E}} \right)^{\alpha} \exp \left[ - (1+\alpha) \frac{E}{\bar{E}} \right] 
\end{equation}
i.e. a superposition of two approximately thermal spectra, with $\bar{p}$ controlling which one dominates. We set $\alpha = 3.5$ for $\bar{\nu}_e$ and $\alpha = 2.5$ for $\bar{\nu}_x$, and $\Gamma$ is a gamma function.

The rates of NS and BH forming collapse events $R_{\mathrm{NS}}(z)$ and $R_{\mathrm{BH}}(z)$ take the form,
\begin{eqnarray} 
R_{\mathrm{NS}}(z) &=& [1-f_{\mathrm{BH}}(z)] R(z) \\
R_{\mathrm{BH}}(z) &=& f_{\mathrm{BH}}(z) R(z)
\end{eqnarray}
\begin{equation}
R(z) = 
\begin{cases}
R_0 (1+z)^{\beta} & \text{for } z \leq z_{\mathrm{th}} \\
R_0 (1+z_{\mathrm{th}})^{\beta} (1+z)^{\gamma} (1+z_{\mathrm{th}})^{-\gamma} & \text{for } z_{\mathrm{th}} < z \leq 4 \\
0 & \text{for } z > 4
\end{cases}
\end{equation}
\begin{equation}
f_{\mathrm{BH}}(z) = 
\begin{cases}
f_0 (1+z)^{\kappa} & \text{for } z \leq z_{\mathrm{th}} \\
f_0 (1+z_{\mathrm{th}})^{\kappa} (1+z)^{\epsilon} (1+z_{\mathrm{th}})^{-\epsilon} & \text{for } z_{\mathrm{th}}  < z \leq 4
\end{cases}
\end{equation}

where $z_{\mathrm{th}}  = 1$ and $\kappa$ and $\epsilon$ are fixed such that $f_{\mathrm{BH}}(z=0) = f_0$, $f_{\mathrm{BH}}(z=1) = f_1$ and $f_{\mathrm{BH}}(z=4) = f_4$ for the parameters $f_0$, $f_1$ and $f_4$. In this case $f_{\mathrm{BH}}(z)$ is the redshift-dependent
fraction of total core-collapse events which result in black holes instead of neutron stars. The form of $R(z)$ is based off an empirical fit to star formation data~\cite{Hopkins:2006bw,Lunardini:2009ya,Yuksel:2012zy}. 
We have made the simplifying assumption that the rate of supernovae and unnovae vanishes for $z > 4$, which is around the redshift where models predict the star formation rate to fall sharply.

Our final step is to calculate the actual measured event rate in a neutrino detector. We focus on electron anti-neutrinos $\bar{\nu}_e$ detected in water Cherenkov detectors such as Super Kamiokande or Hyper-Kamiokande, which  are detected via the observation of positrons from inverse beta decay capture reactions $\bar{\nu}_e + p \rightarrow n + e^+$~\cite{Beacom:2010kk}. The cross section of this interaction $\sigma_{\mathrm{IB}}(E)$ is larger than that for elastic scattering
of the remaining neutrino flavours with electrons, leading  $\bar{\nu}_e$ to dominate the event rate from the DSNB. The detected DSNB spectrum is then,
\begin{equation}
\frac{\mathrm{d}N}{\mathrm{d}E_p} = N_t \Phi(E) \cdot \sigma_{\mathrm{IB}}(E) ,
\end{equation}
where the positron energy $E_p = E - 1.3 \, \mathrm{MeV}$ and $N_t$ is the number of target protons in the detector (with only those in the hydrogen of H$_2$O contributing). Since the Hyper-Kamiokande experiment has a finite energy
resolution $\sigma_E$ we need to account for this when generating our expected spectra. The Hyper-Kamiokande experiment is expected to have the same level of energy resolution compared to Super Kamiokande~\cite{Abe:2011ts}, which takes the form of a gaussian standard deviation~\cite{Abe:2016nxk},
\begin{equation}
\sigma_E(E_p) = \left[ -0.0839 + 0.349 \sqrt{\frac{E_p}{\mathrm{MeV}}} + 0.0397 \left( \frac{E_p}{\mathrm{MeV}} \right) \right] \, \, \mathrm{MeV},
\end{equation}
which is therefore the resolution we adopt for our analysis. We incorporate this to calculate $\frac{\mathrm{d}N}{\mathrm{d}E_{\mathrm{ex}}}$, the measured spectrum expected in Hyper-Kamiokande, using the expression,
\begin{equation}
\frac{\mathrm{d}N}{\mathrm{d}E_{\mathrm{ex}}} = \int \mathrm{d} E_p \frac{\mathrm{d}N}{\mathrm{d}E_p} \frac{1}{\sqrt{2 \pi \sigma_E^2}} \exp \left[ \frac{- (E_{\mathrm{ex}} - E_p)^2}{2 \sigma_E^2} \right] ,
\end{equation}
where $E_{\mathrm{ex}}$ is then the energy measured in Hyper-Kamiokande.

\subsection{Spectra of the DSNB and background events in Hyper-Kamiokande}

\begin{figure}[t]
\centering
\includegraphics[width=0.99\textwidth]{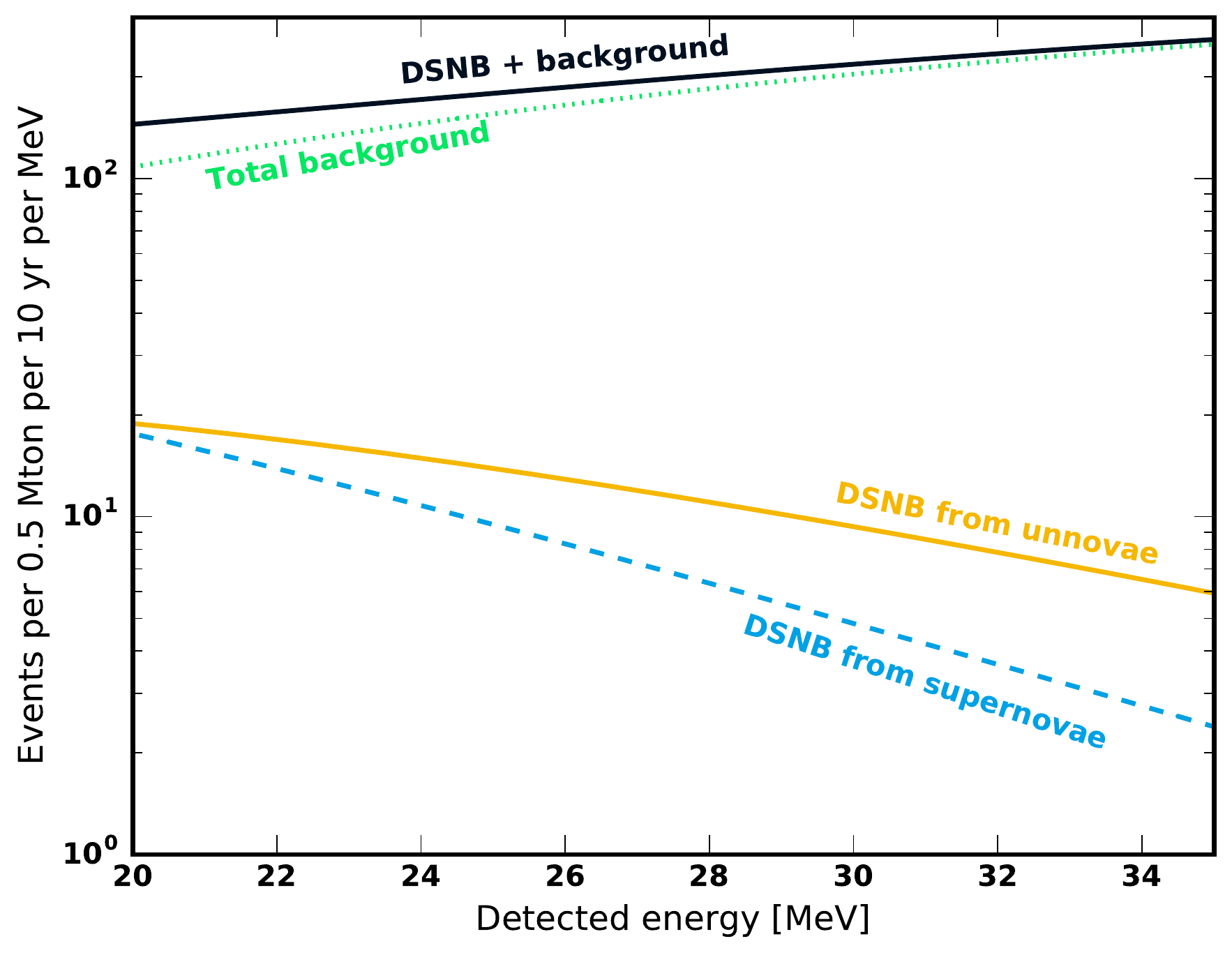}
\caption{Expected spectra of the diffuse neutrino background produced by either NS-forming supernovae (dashed blue) or BH-forming unnovae (solid orange), compared with the background from atmospheric neutrinos and invisible muons
(dotted green). The total spectrum is shown as the solid black line.}
\label{fig:rates_in_detector}
\end{figure}

The dominant background to searches for the DSNB in water Cherenkov detectors above detected energies of 20~MeV comes from atmospheric neutrinos and invisible muons~\cite{Abe:2011ts},
and has a spectrum which rises with energy~\cite{Abe:2011ts,Beacom:2010kk}, which we incorporate into our MCMC study. It has been suggested~\cite{Beacom:2003nk} that doping the water target of Hyper-Kamiokande with gadolinium
could be used to reduce this background, by allowing anti-electron neutrinos to be identified through the tagging of the neutron produced through inverse beta decay. However in this work we do not consider such a possibility,
since it is not clear whether such gadolinium doping will be implemented for the Hyper-Kamiokande experiment.
Below energies of 20~MeV there are expected to be huge backgrounds from reactor neutrinos and the products of spallation reactions~\cite{Abe:2011ts}, and so we set the low-energy threshold of our analysis at 20~MeV i.e. we consider
only $E_{\mathrm{ex}} \geq 20$~MeV.

In figure~\ref{fig:rates_in_detector} we show the expected spectra of the diffuse neutrino background produced by either BH-forming unnovae or NS-forming supernovae, compared with the size of the expected background in Hyper-Kamiokande
above energies of 20~MeV. The number of events expected from the DSNB depends strongly on the many parameters of equation~\ref{eqn:dsnb_all}, and here we use the fiducial values of these parameters outlined in section~\ref{sec:priorvals}.
Fortunately the spectrum from diffuse unnovae neutrinos is expected to be larger than that from supernovae neutrinos, due to the larger average energy predicted by unnovae
simulations~\cite{PhysRevD.78.083014,Fischer:2008rh,Sumiyoshi:2008zw,Lunardini:2009ya}. Even so, there is only a small window between around 20~MeV to 30~MeV where this flux is measurable practically, before it becomes impossible
to distinguish from the background.
Within this window there are expected to be approximately $200$ DSNB events after running Hyper-Kamiokande for 10~years, resulting from a neutrino flux of approximately $\Phi_{\mathrm{BH}} \sim 8 \cdot 10^{-2}$~cm$^{-2}$~MeV$^{-1}$~s$^{-1}$ at $20$~MeV energies from BH-forming unnovae.

\subsection{Choosing the prior distributions for the parameters ~\label{sec:priorvals}}

\begin{table}[t]
\centering
\begin{tabular}{ c || c | c }
\textbf{Parameter} & \textbf{Optimistic priors} & \textbf{Pessimistic priors}  \\ \hline 
$\bar{E}_{e \mathrm{NS}}$& $P \in [14,16]$ MeV  & $P \in [14,16]$  MeV\\
\hline
$\bar{E}_{x \mathrm{NS}}$& $P \in [17,19]$ MeV & $P \in [17,19]$ MeV \\
\hline
$\bar{E}_{e \mathrm{BH}}$& $P \in [23,25]$ MeV & $P \in [15,25]$  MeV\\
\hline
$\bar{E}_{x \mathrm{BH}}$& $P \in [23,28]$ MeV & $P \in [16,33]$  MeV \\
\hline
${L}_{e \mathrm{NS}}$& $P \in [4.5,5.5] \cdot 10^{52} \, \mathrm{ergs}$ &$P \in [4.5,5.5] \cdot 10^{52} \, \mathrm{ergs}$\\
\hline
${L}_{x \mathrm{NS}}$& $P \in [4.5,5.5] \cdot 10^{52} \, \mathrm{ergs}$& $P \in [4.5,5.5] \cdot 10^{52} \, \mathrm{ergs}$ \\
\hline
${L}_{e \mathrm{BH}}$& $P \in [12,14] \cdot 10^{52} \, \mathrm{ergs}$ &$P \in [0,20] \cdot 10^{52} \, \mathrm{ergs}$\\
\hline
${L}_{x \mathrm{BH}}$& $P \in [0.35,0.45] {L}_{e \mathrm{BH}}$ &$P \in [0.3,1] {L}_{e \mathrm{BH}}$ \\
\hline
$R_0$& $P \in [0.8,1.2] \cdot 10^{-4} \mathrm{Mpc}^{-3} \mathrm{s}^{-1}$ & $P \in [0.8,1.2] \cdot 10^{-4} \mathrm{Mpc}^{-3} \mathrm{s}^{-1}$ \\ \hline
$\beta$& $P \propto N(\beta,\mu=3.28,\sigma=0.05)$ & $P \propto N(\beta,\mu=3.28,\sigma=0.05)$ \\ \hline
$\gamma$& $P \propto N(\gamma,\mu=0,\sigma=0.1)$  & $P \propto N(\gamma,\mu=0,\sigma=0.1)$ \\ \hline
$\bar{p}$& $P \in [0.5,0.68]$ & $P \in [0,0.68]$ \\ \hline
$f_0$& $P \in [0,1]$  & $P \in [0,1]$ \\ \hline
$f_1$& $P \in [0,1]$  & $P \in [0,1]$ \\ \hline
$f_4$& $P \in [0,1]$  & $P \in [0,1]$ 
\end{tabular}
\caption{Priors for each of our parameters in either the optimistic or pessimistic case. Priors are flat within the range and zero outside unless otherwise stated, and $N(x,\mu,\sigma) = \frac{1}{\sqrt{2 \pi \sigma^2}} \exp \left[ \frac{- (x - \mu)^2}{2 \sigma^2} \right]$ represents a normal distribution with mean $\mu$ and standard deviation $\sigma$.}
\label{priors}
\end{table}
We are interested primarily in $R_{\mathrm{BH}}(z)$ while the remaining parameters are essentially nuisance parameters. Due to the large number of variables we perform a MCMC over our 15
parameters listed in table~\ref{table_params} by comparing theoretical predictions from equation~\ref{eqn:dsnb_all} to simulated data. For the simulated data we take fiducial values of 
$\bar{E}_{e \mathrm{NS}} = 15$~MeV, $\bar{E}_{x \mathrm{NS}} = 18$~MeV, $\bar{E}_{e \mathrm{BH}} = 23.6$~MeV, $\bar{E}_{x \mathrm{BH}} = 24.1$~MeV, ${L}_{e \mathrm{NS}} = 5 \cdot 10^{52} \, \mathrm{ergs}$,
${L}_{x \mathrm{NS}} = 5 \cdot 10^{52} \, \mathrm{ergs}$, ${L}_{e \mathrm{BH}} = 12.8 \cdot 10^{52} \, \mathrm{ergs}$, ${L}_{x \mathrm{BH}} = 4.9 \cdot 10^{52} \, \mathrm{ergs}$,
$R_0 = 10^{-4} \mathrm{Mpc}^{-3} \mathrm{s}^{-1}$, $\beta = 3.28$, $\gamma = 0$, $\bar{p} = 0.68$ and $f_0 = f_1 = f_4 = 0.2$. We assume a 500 kilo-tonne water Cherenkov experiment similar to Hyper-Kamiokande with 10 years worth of data.
The simulated data is generated by sampling a discrete set of events randomly from the total theoretical spectrum (including the background) according to Poisson statistics, then binning them into a histogram. This means that the data-set will include fluctuations from the ``true'' spectrum which one would expect from real experimental data.
The simulated data-set is then essentially one example of what Hyper Kamiokande might see, given the fiducial parameter values chosen here.
Since these fluctuations are by-their-nature random, our results could in principle depend on the simulated data-set. To test this we have cross-checked our analysis with ten different simulated datasets, and find very similar posterior
contours in each case, and so it is reasonable to assume that the effect of the simulated data-set itself on our results is negligable.

Our choice of priors on our parameters is important, especially for the case of the BH-forming unnovae where we only have information from simulations~\cite{PhysRevD.78.083014,Fischer:2008rh,Sumiyoshi:2008zw,Lunardini:2009ya}.
In order to fully understand the effect of prior choice we consider two different scenarios. In the first scenario, labelled as ``optimistic'' we assume we have good
knowledge of the luminosity and spectra of both NS-forming supernovae and BH-forming unnovae, while for the second case labelled ``pessimistic'' we assume poor knowledge of such spectra. In both cases we assume no prior knowledge of
$R_{\mathrm{BH}}(z)$. The full prior list is given in table~\ref{priors}. 

In the optimistic case we assume good knowledge of the average energies and luminosities of all neutrino flavours associated with unnovae, based on the results of simulations~\cite{PhysRevD.78.083014,Lunardini:2009ya}. However the pessimistic
case differs in that we take broad priors on these quantities associated with unnovae, which include parameter values for which the spectra of neutrinos from BH-forming unnovae are identical to those for NS-forming supernovae
or where ${L}_{e \mathrm{BH}}=0$ and so unnovae do not produce neutrinos in vast amounts as supernovae do. In 
this case the assumption is that the simulations are wrong and that there is no difference between the neutrino emission in either case, and so we can learn little to nothing about black holes from the DSNB.
The pessimistic case also differs in our choice for the prior on $\bar{p}$, where we assume that oscillation effects within the supernovae or unnovae are not well understood, while for the optimistic case we assume that
we will have a high-statistics measurement of  $\bar{p}$ by the time precision measurements of the DSNB are made.
For the remaining parameters we assume that by the time Hyper-Kamiokande has enough data to make a precision study of the DSNB, we will have accurate knowledge of the parameters associated with neutrino emission from NS-forming
supernovae and of $R(z)$~\cite{Lien:2010yb}. In the former case this may be because a galactic supernova has occurred by this time, and its neutrino emission has been observed to high accuracy. 

We have assumed flat priors on $f_0$, $f_1$ and $f_4$, however in principle it should be possible to constrain the fractional function $f_{\mathrm{BH}}(z)$ using either direct measurements of the supernova rate as a function of redshift, or observations of the rate of stars which disappear, which may be related
to the rate of unnovae~\cite{Yuksel:2012zy,Kochanek:2008mp,Lien:2010yb,Adams:2016ffj,Adams:2016hit}. By the time the DSNB has been measured to high precision, searches for disappearing stars close to our own galaxy such as in refs.~\cite{Adams:2016ffj,Adams:2016hit} may be 
advanced enough to give us prior information on $f_0$. Hence although
we assume flat priors on $f_0$, $f_1$ and $f_4$ this may not be appropriate for future studies when more data may be available. By combining the DSNB with information from other surveys our projections can only improve, and so
our study can be considered as a worst-case-scenario where only information from the DSNB is available for the black hole birth rate.
The prior range for $R_0$ is chosen based on the quoted uncertainty at $z = 0$ on the cosmic star formation rate from ref.~\cite{Lien:2010yb}. By the time the DSNB is measured to high-precision, it is likely that more advanced synoptic
surveys will have been performed, reducing the size of the uncertainty on $R_0$~\cite{Lien:2010yb}.

\subsection{Results of the MCMC projection for the DSNB \label{sec:results_mcmc}}
\begin{figure}[t]
\centering
\includegraphics[width=0.47\textwidth]{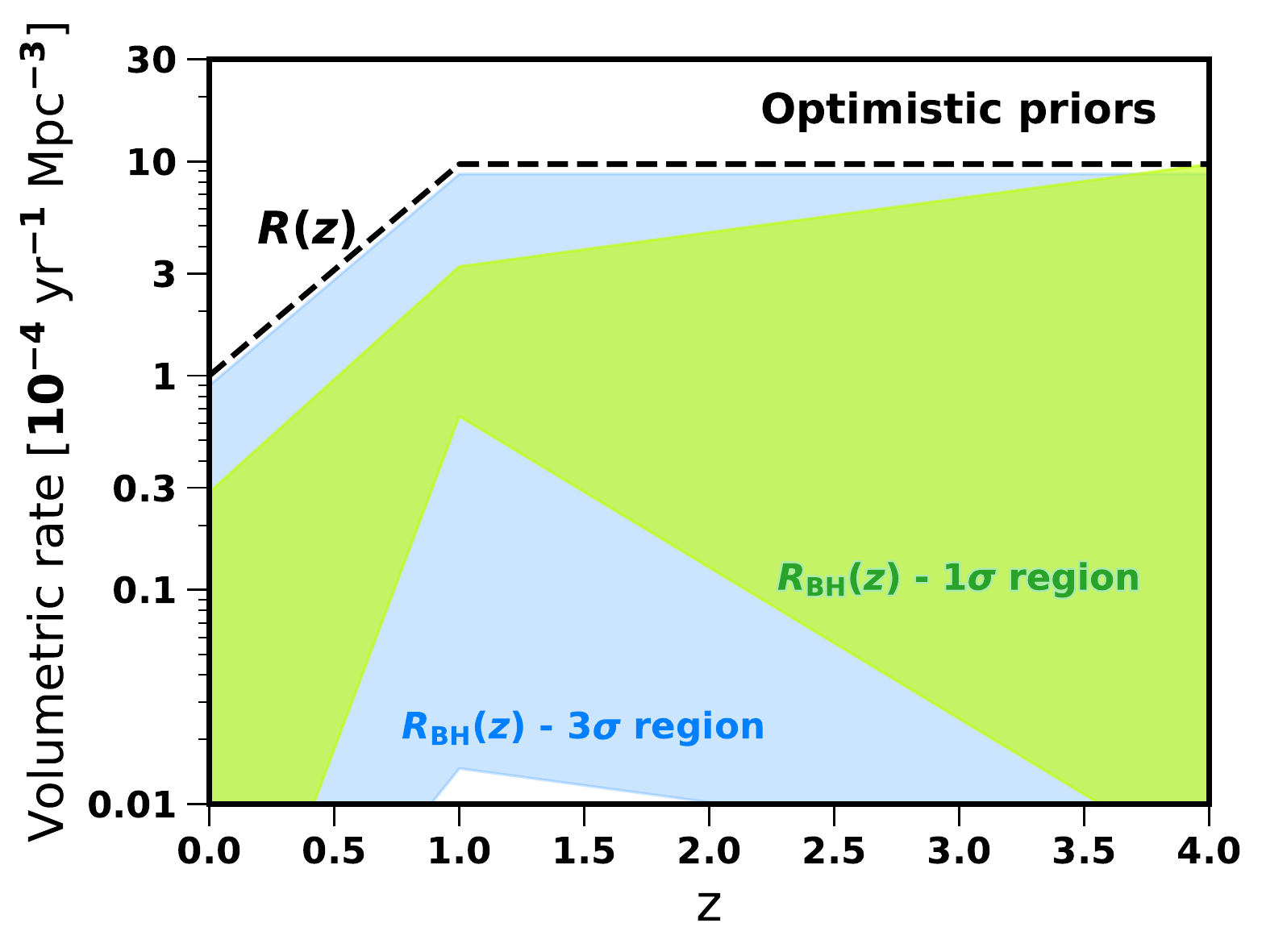} \hspace{10pt}
\includegraphics[width=0.47\textwidth]{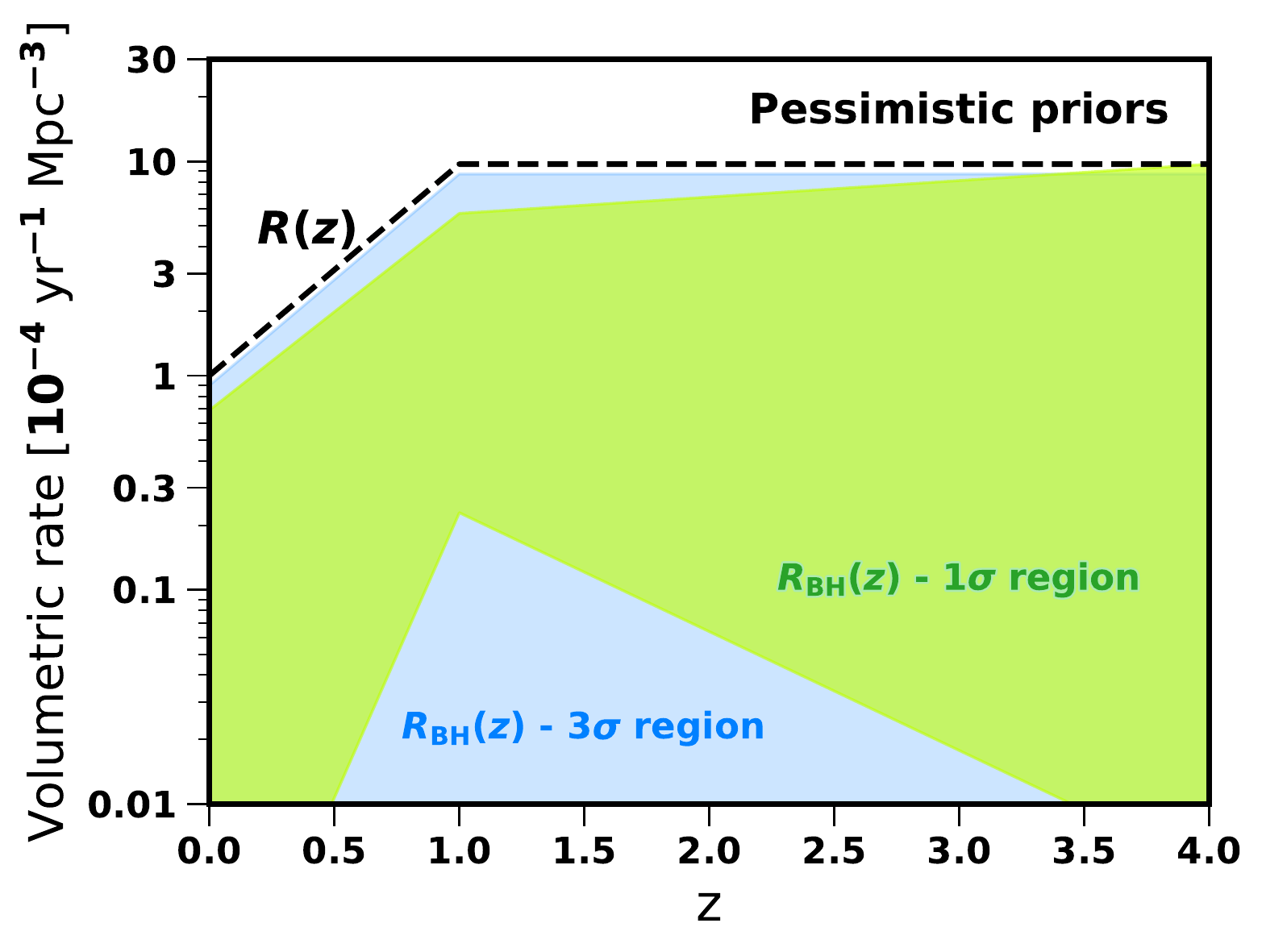}
\caption{\textbf{Left:} One sigma (green) and three sigma (blue) regions for the birth rate of black holes $R_{\mathrm{BH}}(z)$ assuming optimistic priors, inferred from the DSNB with 10 years worth of data in an experiment similar to Hyper-Kamiokande. \textbf{Right:} Same but for the pessimistic case.}
\label{fig:rates_from_dsnb}
\end{figure}

The results of our scan over the parameters described in the previous section are shown in figure~\ref{fig:rates_from_dsnb} and in the corner plots of figures~\ref{fig:corner_op} and \ref{fig:corner_pess}, which are described in more detail in appendix~\ref{app:MCMC}. As is clear from figure~\ref{fig:rates_from_dsnb},  in the optimistic case
we know the unnovae neutrino spectrum well enough to infer strong bounds on the black hole birth rate $R_{\mathrm{BH}}(z)$  at least up to redshift $z = 1$. In contrast for the pessimistic case the constraints are much weaker at $95\%$ confidence,
since we no longer claim to know the unnovae neutrino spectrum. 

The difference between the priors can be understood from the corner plots of figures~\ref{fig:corner_op} and \ref{fig:corner_pess}. 
As can be seen in figure~\ref{fig:corner_pess},
there is a degeneracy between $f_0$ or $f_1$ and $\bar{E}_{e \mathrm{BH}}$ and between $f_1$ and ${L}_{e \mathrm{BH}}$, which weakens the bounds on $f_0$ and $f_1$ in the pessimistic case. For example if for the pessimistic case ${L}_{e \mathrm{BH}}$
can be close to zero then $f_1$ needs to be much larger as a result. While for the optimistic case such values of ${L}_{e \mathrm{BH}}$ and $\bar{E}_{e \mathrm{BH}}$ are not allowed as they contradict results form simulations of
BH-forming collapse events, and so the bounds on $f_0$ and $f_1$ are much tighter.
This degeneracy is also the reason as to why the posterior for $\bar{E}_{e \mathrm{BH}}$ is not centred on its fiducial value, as it is being affected by the potential for $f_0$ or $f_1$ to take values above $0.2$. Specifically, as shown
in the two-dimensional plots of  $f_0$ or $f_1$ vs. $\bar{E}_{e \mathrm{BH}}$,
the larger $f_0$ or $f_1$ gets the smaller $\bar{E}_{e \mathrm{BH}}$ needs to be such that the unnova spectrum still provides a good fit to the simulated data, since both parameters change the total expected flux of unnovae neutrinos.
Hence since the one-dimensional plot of $\bar{E}_{e \mathrm{BH}}$ is an integral of any of the two-dimensional plots over the other parameter (e.g. $f_1$ or $f_0$) the peak of the distribution is shifted to lower values.
If $f_0$ and $f_1$ were fixed at their fiducual values, then the posterior distribution of $\bar{E}_{e \mathrm{BH}}$ would 
be centred on its fiducial value.
Indeed $f_0$ and $f_1$ themselves have some degeneracy between each other, however this is partly broken since the energy in equation~(\ref{eqn:dsnb_all}) is redshifted, meaning that the spectrum of neutrinos from $z=1$ is different from
those produced near redshift $z = 0$.

There is also a degeneracy between $\bar{p}$ and $f_0$ which weakens the bounds on the BH birth rate at low redshift values, particularly for the pessimistic prior set. This arises from the fact that a value of $\bar{p}$ close to zero means that the neutrino spectra from both NS
and BH forming collapse events are harder, which mimics the effect on the tail of the DSNB caused by having a different value of $f_0$.
These are the strongest degeneracies between our parameters and $f(z)$, which is why we do not show all of the parameters in figures~\ref{fig:corner_op} and \ref{fig:corner_pess}.

For both sets of priors $f_4$ is poorly constrained, and therefore so is $R_{\mathrm{BH}}(z)$ approaching $z = 4$. This can be seen in figures~\ref{fig:corner_op} and \ref{fig:corner_pess}, for example in the two-dimensional 
posterior plots where the contours vary only a small amount with a changing $f_4$, and in the one-dimensional plot where the posterior value changes little between $f_4=0$ and $f_4=1$, as compared with $f_0$ or $f_1$.
This poor constraint is due to several factors: it partly results from the suppression of the rate at larger redshifts by the cosmological factor
multiplying $\Omega_m$, and partly also because the energies of the neutrinos from $z = 4$ have been redshifted to much smaller values where the cross section for detection $\sigma_{\mathrm{IB}}(E)$ is smaller, and where the supernovae and unnovae components
are more difficult to separate.

Our MCMC shows that, as expected, the precision to which we can infer $R_{\mathrm{BH}}(z)$ from the DSNB depends crucially on how well we understand the spectrum of neutrinos from BH-forming unnovae. If we trust simulations of unnovae~\cite{PhysRevD.78.083014,Fischer:2008rh,Sumiyoshi:2008zw,Lunardini:2009ya} then  ${L}_{e \mathrm{BH}}$ and $\bar{E}_{e \mathrm{BH}}$ are known well enough to fix $R_{\mathrm{BH}}(z)$. However in the pessimistic case, $R_{\mathrm{BH}}(z)$ is only weakly constrained, since we are unable to exploit even a high-precision measurement of the DSNB in an experiment like Hyper-Kamiokande.

\section{Combining posteriors from the MCMC with data from LIGO \label{sec:ligo}}

After the detection of gravitational waves from mergers of BH-BH binaries by LIGO~\cite{TheLIGOScientific:2016htt,Abbott:2016blz}, we are entering a period where the merger rate of black holes can be measured with potentially unprecedented precision~\cite{Elbert:2017sbr,Kovetz:2016kpi}.
The current best-fit value for the BH-BH merger rate from LIGO  $\mathcal{R}_{\mathrm{BH-BH}}$ is within the range $9 - 240 \, \mathrm{Gpc}^{-3} \mathrm{yr}^{-1}$ at $90\%$ confidence~\cite{TheLIGOScientific:2016pea}.
In the previous section we saw that the birth rate of black holes $R_{\mathrm{BH}}(z)$ can be inferred from the DSNB with future neutrino experiments such as Hyper-Kamiokande. This function is related to the merger rate of black holes at the present time $t_0$ by the following equation,
\begin{equation}
 \mathcal{R}_{\mathrm{BH-BH}} =  \frac{\epsilon}{2} \int_0^{t_0} \mathrm{d} t \, R_{\mathrm{BH}}(t_0 - t) P(t),
\label{eqn:R_eq_1}
\end{equation}
where we have written $R_{\mathrm{BH}}(z)$ as a function of time $t$ and $P(t)$ is the distribution of expected merger times for BH-BH binary pairs,
which can be obtained from simulations of binary mergers~\cite{Eldridge:2016ymr,Mandel:2015qlu,deMink:2016vkw,Elbert:2017sbr,Dominik:2012kk}. The latter usually takes a form close to a Normal distribution centred on an average merger
time around several Gyr.
The factor $\epsilon$ is the merger fraction, which equals the fraction of black holes which lead to merger events at the present day i.e. the ratio of the merger rate density today to the density of black holes available to merge.
This will be several orders
of magnitude smaller than unity due to e.g. the small fraction of black holes expected to be in binary systems.
In ref.~\cite{Elbert:2017sbr} it was shown that with a theoretical prediction for $R_{\mathrm{BH}}(t)$ one can use $\mathcal{R}_{\mathrm{BH-BH}}$ to place constraints on the unknown quantity $\epsilon$. 
Although this is perfectly reasonable, there is no reason \emph{a priori} to assume any particular redshift dependence for the black hole birth rate, for example.
Hence instead we focus on a data-driven approach and use $R_{\mathrm{BH}}(z)$ from our MCMC study of the DSNB. 

\begin{figure}[t]
\centering
\includegraphics[width=0.47\textwidth]{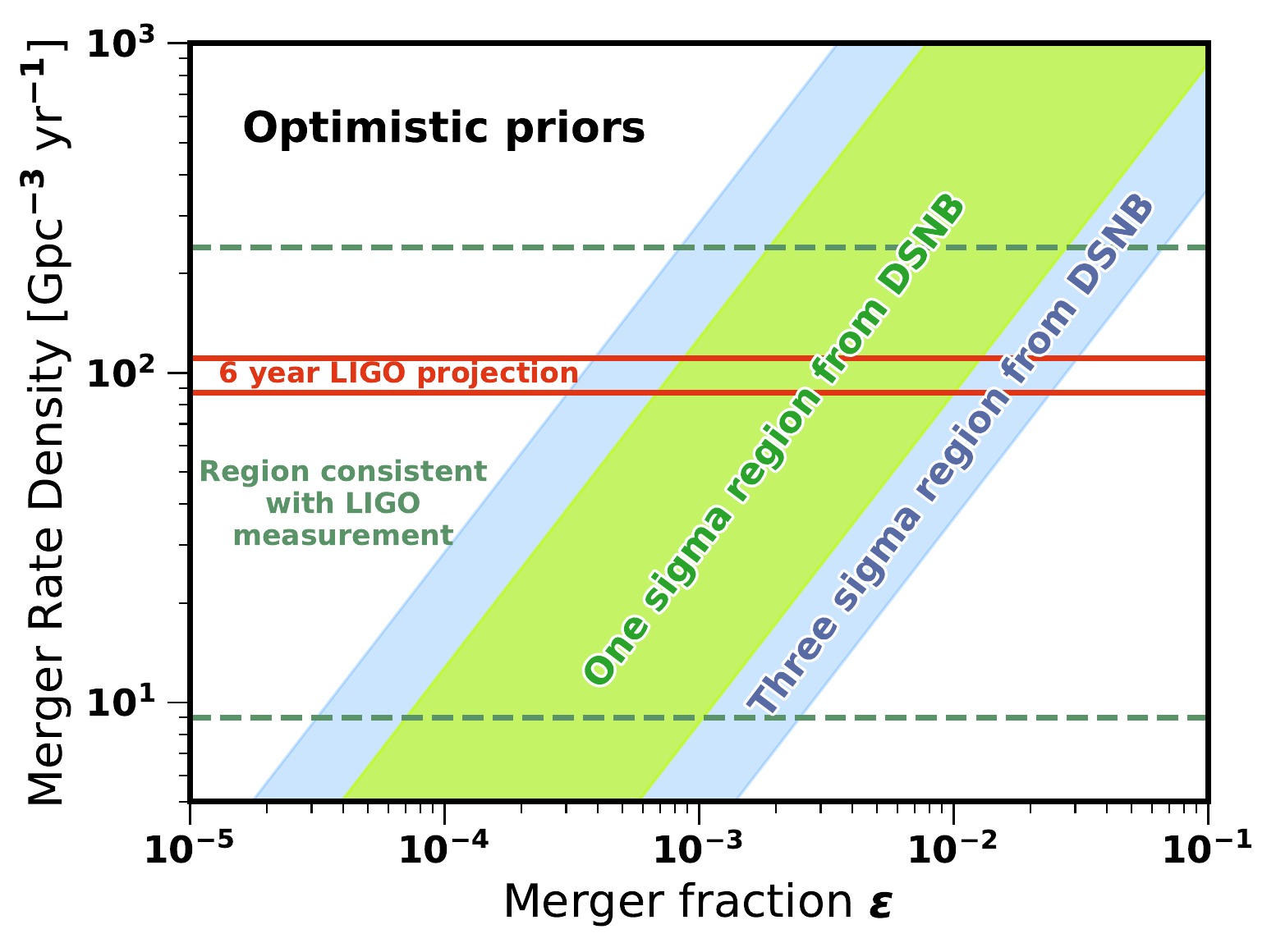} \hspace{10pt}
\includegraphics[width=0.47\textwidth]{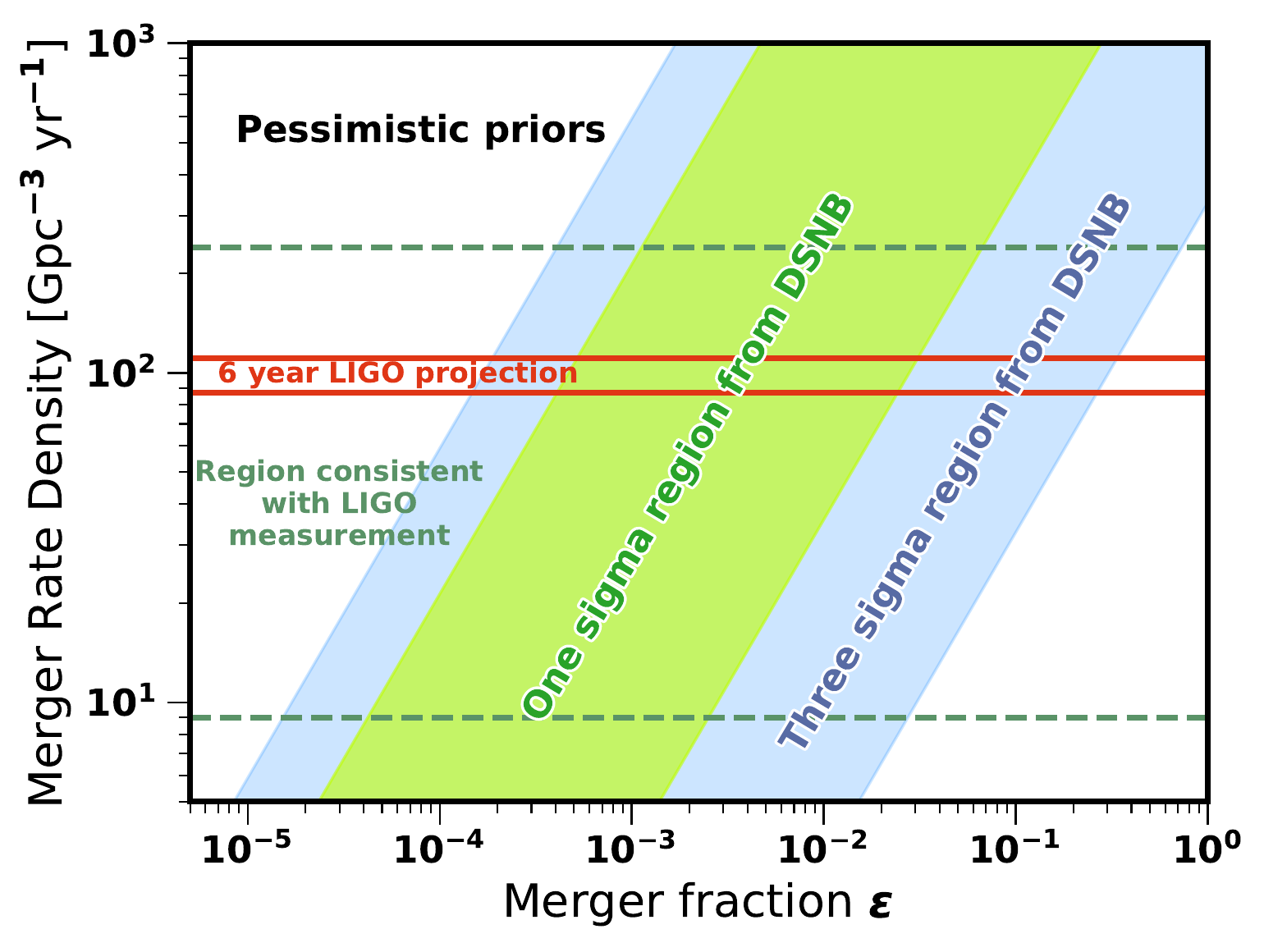}
\caption{The merger rate of black hole binaries at the present time calculated using equation~(\ref{eqn:R_eq_1}) versus the merger fraction $\epsilon$, using the merger time distribution from simulations~\cite{Mandel:2015qlu,deMink:2016vkw}. The horizontal lines show
upper and lower limits on the merger rate density from either LIGO measurements (green)~\cite{TheLIGOScientific:2016pea,Abbott:2016blz,TheLIGOScientific:2016htt,Elbert:2017sbr} or projections for LIGO running after six years (red)~\cite{Kovetz:2016kpi}.
The filled blue region shows all merger rates consistent with the three sigma interval of $R_{\mathrm{BH}}(z)$, and the filled light green region is the one-sigma interval.}
\label{fig:merger_frac_with_sim_func}
\end{figure}

In order to infer the black hole merger rate from the birth rate $R_{\mathrm{BH}}(z)$ we need to know the distribution of merger time-scales $P(t)$. Here we use results from simulations~\cite{Eldridge:2016ymr,Mandel:2015qlu,deMink:2016vkw,Elbert:2017sbr,Dominik:2012kk}
to estimate this function, which has the merger time distributed between $4$~Gyr and $10$~Gyr with a maximum of the probability distribution at $6$~Gyr (where $z \approx 0.6$). This leads to our results shown in figure~\ref{fig:merger_frac_with_sim_func}. The filled blue/green region shows all present-day BH-BH merger rates consistent with the three/one sigma region of $R_{\mathrm{BH}}(z)$ from figure~\ref{fig:rates_from_dsnb}.
Where this region intersects with the upper and lower bounds from LIGO tells us the approximate bounds on the merger fraction $\epsilon$.
Since the LIGO collaboration provide the full posterior for their inferred BH-BH merger rate in ref.~\cite{TheLIGOScientific:2016pea} we combine this with our posterior from the MCMC on the BH birth rate as a function of redshift
to obtain statistically robust predictions for $\epsilon$. We do this by scanning over $\mathcal{R}_{\mathrm{BH-BH}}$ and $R_{\mathrm{BH}}(z)$, each time calculating the value of $\epsilon$ according
to equation~(\ref{eqn:R_eq_1}), and calculating the combined posterior by multiplying the posterior for $\mathcal{R}_{\mathrm{BH-BH}}$ from the LIGO collaboration with our own for the MCMC. For each value of $\epsilon$ the maximum
value of this combined posterior gives the resulting posterior distribution for $\epsilon$ which we then use to calculate confidence intervals.
With the current upper and lower bounds on the merger rate from LIGO we are able to set a limit on the merger fraction at the level of $3 \cdot 10^{-5} \leq \epsilon \leq 5 \cdot 10^{-2}$ for the optimistic prior set and 
$1.5 \cdot 10^{-5} \leq \epsilon \leq 5 \cdot 10^{-1}$ for the pessimistic prior set, both at the $3 \sigma$ level.

Given that our DSNB study assumes an experiment similar to Hyper-Kamiokande running for 10 years, a scenario which is not currently practical, we consider the possibility that by the time neutrino experiments have collected
enough data, gravitational wave experiments should have a significantly better measurement of $\mathcal{R}_{\mathrm{BH-BH}}$. For example in ref.~\cite{Kovetz:2016kpi} it was shown that after six years of running, LIGO could be able to measure
$\mathcal{R}_{\mathrm{BH-BH}}$ to a precision of $11.85$~Gpc$^{-3}$yr$^{-1}$ at one-sigma. Assuming that this results in a Gaussian posterior on $\mathcal{R}_{\mathrm{BH-BH}}$ with a standard deviation of $11.85$~Gpc$^{-3}$yr$^{-1}$ we derive 
confidence intervals for $\epsilon$ in the same way as for the current LIGO result.
As can be seen from figure~\ref{fig:merger_frac_with_sim_func} this allows the constraints on $\epsilon$ to be tightened significantly, to the range $2 \cdot 10^{-4} \leq \epsilon \leq 3 \cdot 10^{-2}$ at $3 \sigma$
confidence for the optimistic prior set and $1 \cdot 10^{-4} \leq \epsilon \leq 2 \cdot 10^{-1}$ at $3 \sigma$ confidence for the pessimistic prior set.

\begin{figure}[t]
\centering
\includegraphics[width=0.47\textwidth]{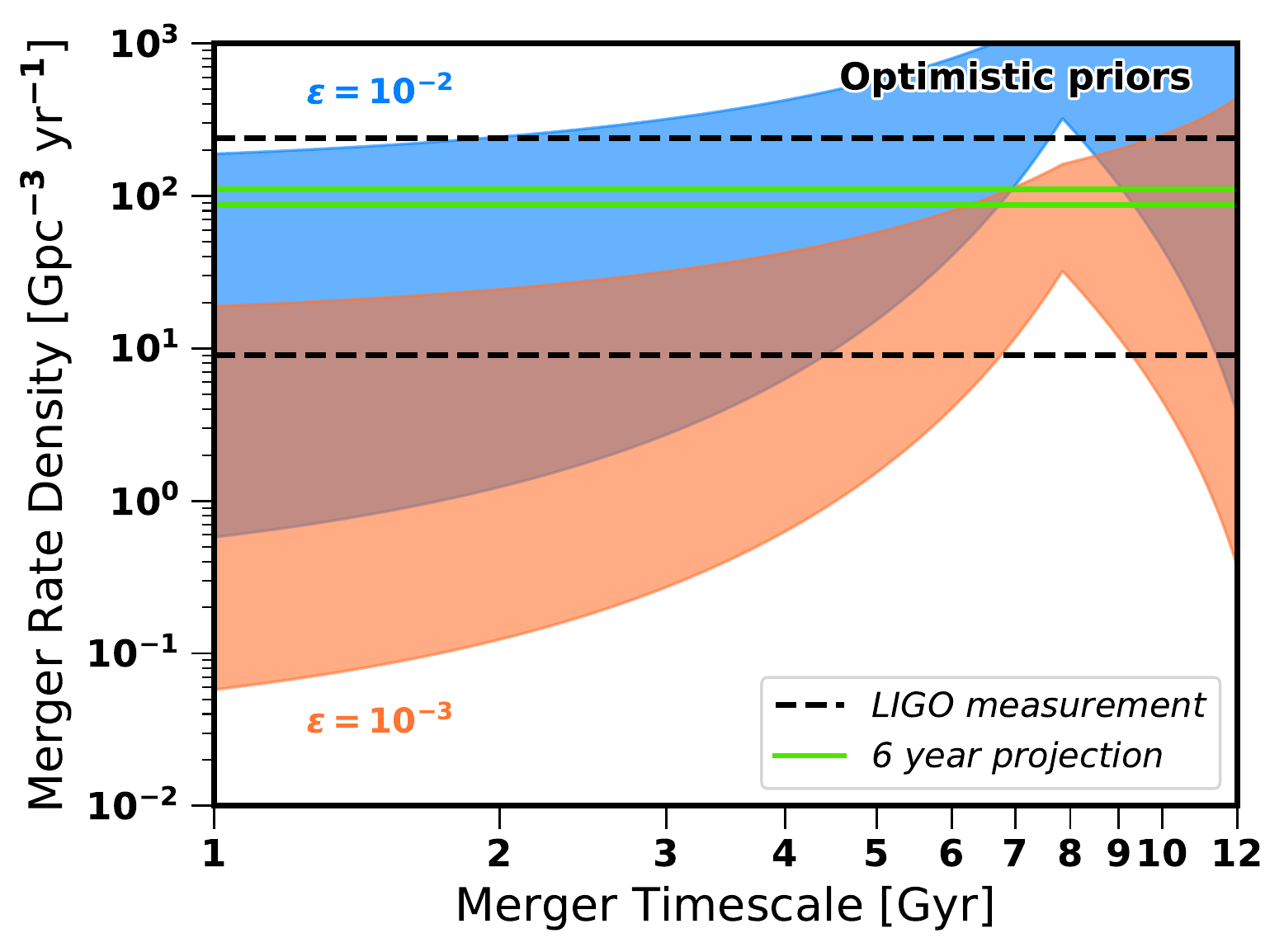} \hspace{10pt}
\includegraphics[width=0.47\textwidth]{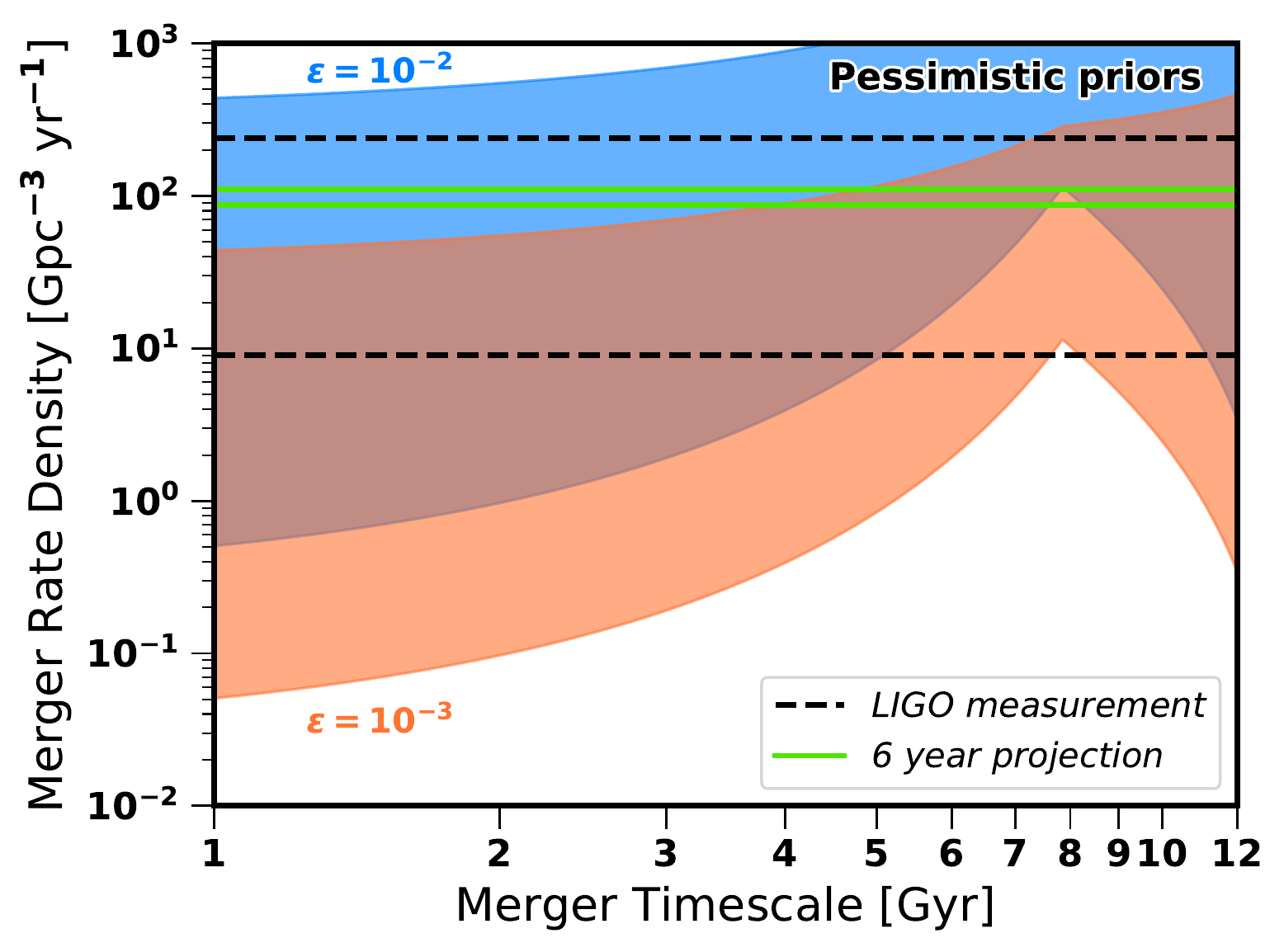}
\caption{Similar to figure~\ref{fig:merger_frac_with_sim_func} but for the case where the black hole binary merger timescale is unknown. The merger rate of black hole binaries calculated from equation~(\ref{eqn:R_eq_2}) for two different values of the merger fraction $\epsilon$, assuming that all black hole binaries have the same unknown merger time-scale. Dashed horizontal lines show
upper and lower limits on the merger rate density from either LIGO measurements (dashed)~\cite{TheLIGOScientific:2016pea,Abbott:2016blz,TheLIGOScientific:2016htt,Elbert:2017sbr} or projections for LIGO running after six years (solid)~\cite{Kovetz:2016kpi}.}
\label{fig:rate_vs_timescale}
\end{figure}

Alternatively if we have no prior knowledge of the merger time-scale we can follow ref.~\cite{Elbert:2017sbr} in making the simplifying assumption that $P(t)= \delta(t - \tau)$ in equation~(\ref{eqn:R_eq_1}) leading to the simplified expression,
\begin{equation}
 \mathcal{R}_{\mathrm{BH-BH}} =  \frac{\epsilon}{2} R_{\mathrm{BH}}(t_0 -\tau),
\label{eqn:R_eq_2}
\end{equation}
where now $\tau$ is the (unknown) merger time of the binary system.
In this case we use equation~(\ref{eqn:R_eq_2}) to combine the bounds on the merger rate $ \mathcal{R}_{\mathrm{BH-BH}}$ from LIGO with our constraint on the black hole birth rate from our MCMC study in the previous section. Shown in figure~\ref{fig:rate_vs_timescale}
are the allowed regions for the black hole merger rate as a function of merger time-scale $\tau$ for two different values of $\epsilon$, from our MCMC, compared with the LIGO measurement of the BH-BH merger rate.
If the merger timescale is unknown then any value of $\epsilon$ which yields the correct merger rate density today, as measured by LIGO, for any value of the merger timescale is allowed.
For the optimistic prior set our projection is that $\epsilon$ can be constrained within the range $2 \cdot 10^{-5} \leq \epsilon \leq 5 \cdot 10^{-1}$ using the current LIGO merger rate and $2 \cdot 10^{-5} \leq \epsilon \leq 2 \cdot 10^{-1}$ for the 6-year projection,
while for the pessimistic prior the bounds are projected to be $3 \cdot 10^{-5} \leq \epsilon \leq 6 \cdot 10^{-1}$ using the LIGO measured rate and $3 \cdot 10^{-5} \leq \epsilon \leq 3 \cdot 10^{-1}$ for the 6-year projection, all at the $1 \sigma$ level. Without knowing the 
merger time-scale distribution these constraints are poor, and at any higher level of confidence $\epsilon$ is essentially unconstrained.

Comparing figures~\ref{fig:merger_frac_with_sim_func} and \ref{fig:rate_vs_timescale} it is clear that knowledge of the merger timescale distribution of black hole binaries $P(t)$ vastly reduces the uncertainty on $\epsilon$ for both sets of priors.
Hence in order to place strong constraints on $\epsilon$ the merger time-scale for black hole binaries needs to be known.
Our constraints on the merger fraction $\epsilon$ could be improved upon in several ways:
the first is if the LIGO collaboration were able to constrain the merger time-scale of each event, or if we otherwise knew this time-scale to better precision. The second is through additional information on the black hole birth
rate e.g. through precision measurements of the supernova or unnova rate~\cite{Yuksel:2012zy,Kochanek:2008mp,Lien:2010yb}.
A third method would be through theoretical predictions or models of the black hole birth rate, which could complement data from the DSNB.
Indeed a disadvantage of our approach is that we are not able to make predictions for the black hole birth rate in individual galaxies, and so we can not exploit the directional sensitivity of LIGO to BH-BH mergers, but theoretical
models could help with this~\cite{Elbert:2017sbr}. 
In addition we have made the assumption that the neutrino spectrum and flux varies only between either NS or BH forming collapse events, but not within these categories. This seems to be a plausible assumption based on results
from simulations~\cite{PhysRevD.78.083014}, but may not be physical, and could be incorporated into a more advanced MCMC analysis to improve precision. 

\section{Conclusion \label{sec:conc}}

In this work we have shown that neutrinos and gravitational waves present complementary probes of black hole physics~\cite{Lunardini:2009ya,Yuksel:2012zy,Kovetz:2016kpi,Elbert:2017sbr}, since both can travel cosmological distances without being significantly
perturbed~\cite{Baker:2016reh}. We found that neutrino experiments such as Hyper-Kamiokande will become effective probes of black hole physics through precision measurements of the DSNB,
allowing the black hole birth rate to be determined to a precision which is not possible otherwise.
When combined with data from experiments looking for gravitational waves, such as LIGO~\cite{Abbott:2016blz,TheLIGOScientific:2016htt}, the  black hole merger fraction can also be measured.

We have performed a MCMC projection for measuring the black hole birth rate $R_{\mathrm{BH}}(z)$ from the DSNB, detected in an experiment like Hyper-Kamiokande running for 10 years.
The high-energy tail of the DSNB should contain neutrinos from BH-forming core collapse events, known as unnovae,  and so the size and spectral shape of this tail is an effective probe of the BH birth rate. This is shown in figure~\ref{fig:rates_in_detector}.
However there are as many as $15$ parameters involved in this calculation (see table~\ref{table_params}), and this work is the first to take all of these fully into account.
Since our only knowledge of the BH-forming unnova neutrino spectrum comes from simulations~\cite{PhysRevD.78.083014,Fischer:2008rh,Sumiyoshi:2008zw,Lunardini:2009ya}, we have performed two MCMC analyses with different sets of priors, shown in table~\ref{priors}.
Our optimistic prior set worked on the assumption that the unnova spectrum is well-understood, leading to strong constraints on $R_{\mathrm{BH}}(z)$, as shown in figure~\ref{fig:rates_from_dsnb}.
However our pessimistic set led to much weaker constraints on the black hole birth rate, since it had wide priors on the unnova neutrino spectrum and flux.

By combining our posterior distributions for $R_{\mathrm{BH}}(z)$ with data from the LIGO experiment~\cite{Abbott:2016blz,TheLIGOScientific:2016htt}
and projections for its future precision, on the BH-BH merger rate, we calculated projected constraints on $\epsilon$, the (unknown) ratio of the observed merger rate density to the density of black holes which were born long enough ago to be available to result in merger events today, as can be seen in figures~\ref{fig:merger_frac_with_sim_func} and~\ref{fig:rate_vs_timescale}. 
For the optimistic prior set our best projected constraint after ten years of LIGO and Hyper-Kamiokande data is at the level of $2 \cdot 10^{-4} \leq \epsilon \leq 3 \cdot 10^{-2}$ at $3 \sigma$ confidence and for the
pessimistic priors it is $1 \cdot 10^{-4} \leq \epsilon \leq 2 \cdot 10^{-1}$ at $3 \sigma$ confidence, for the case of figure~\ref{fig:merger_frac_with_sim_func} where the BH binary merger time-scale is known.

We note also that a measurement of the black hole birth rate from the DSNB on its own may also provide information on 
the formation of black holes e.g. on the progenitor masses of stars which form into black holes, though this requires further study.
It also provides complementary information on $f(z)$ to searches in the optical spectrum for disappearing stars~\cite{Adams:2016ffj,Adams:2016hit}.

\acknowledgments
We thank Thomas Dent and Christopher Kochanek for comments on the manuscript and Ilya Mandel for useful comments on the merger rate of black holes.
The research leading to these results has received funding from the European Research Council through the project DARKHORIZONS under the European Union's Horizon 2020 program (ERC Grant Agreement no.648680).  The work of MF was supported partly by the STFC Grant ST/L000326/1.

\appendix
\section{Corner plots from the MCMC analysis \label{app:MCMC}}
\begin{figure}[t]
\includegraphics[width=0.97\textwidth]{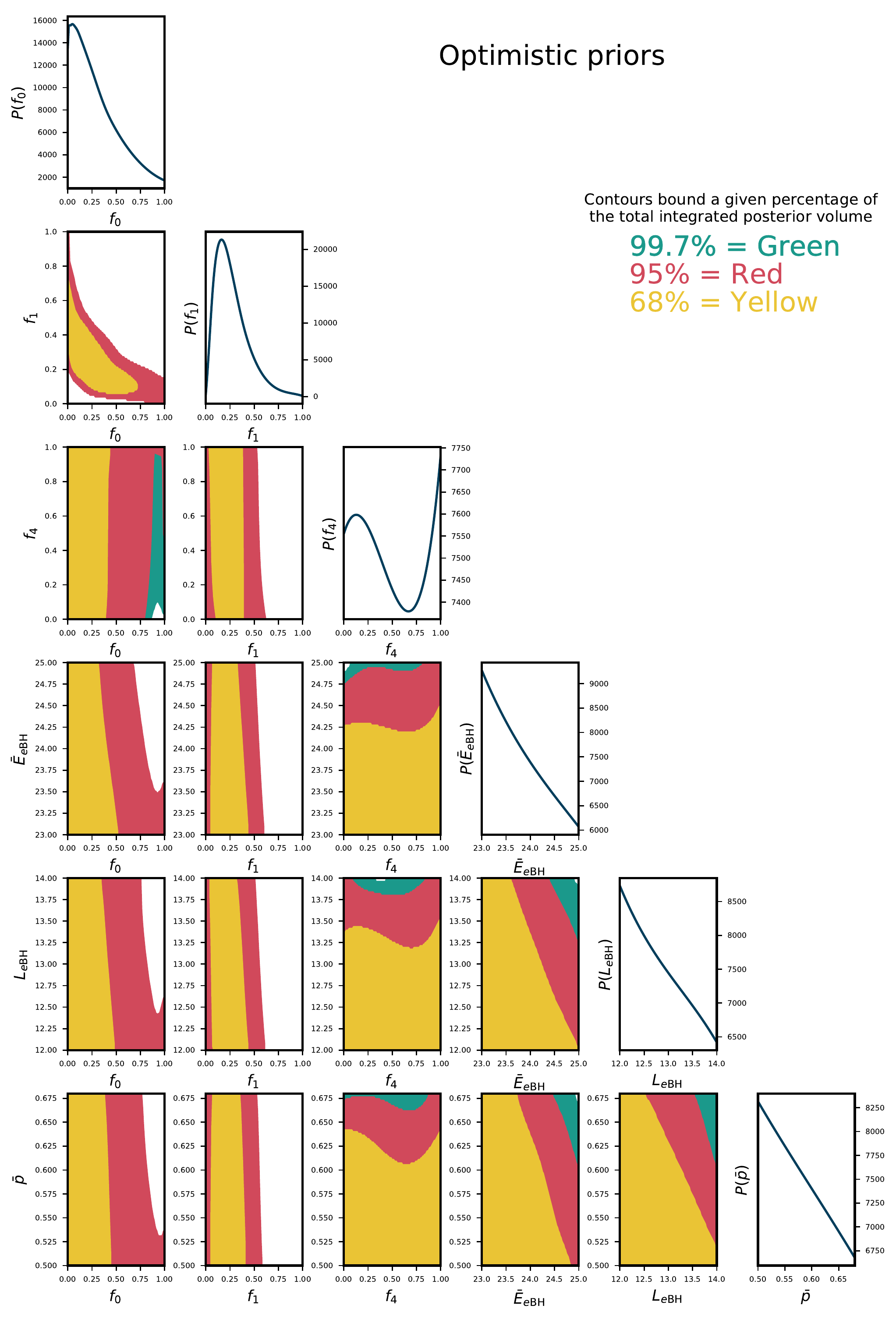}
\caption{Corner plot of of five parameters from our MCMC projection from running an experiment similar to Hyper-Kamiokande for 10 years with optimistic priors. The contours bound a given fraction of the total integrated posterior.
For the one-dimensional histograms, $P(x)$ is the one-dimensional posterior for the parameter $x$.}
\label{fig:corner_op}
\end{figure}

\begin{figure}[t]
\includegraphics[width=0.97\textwidth]{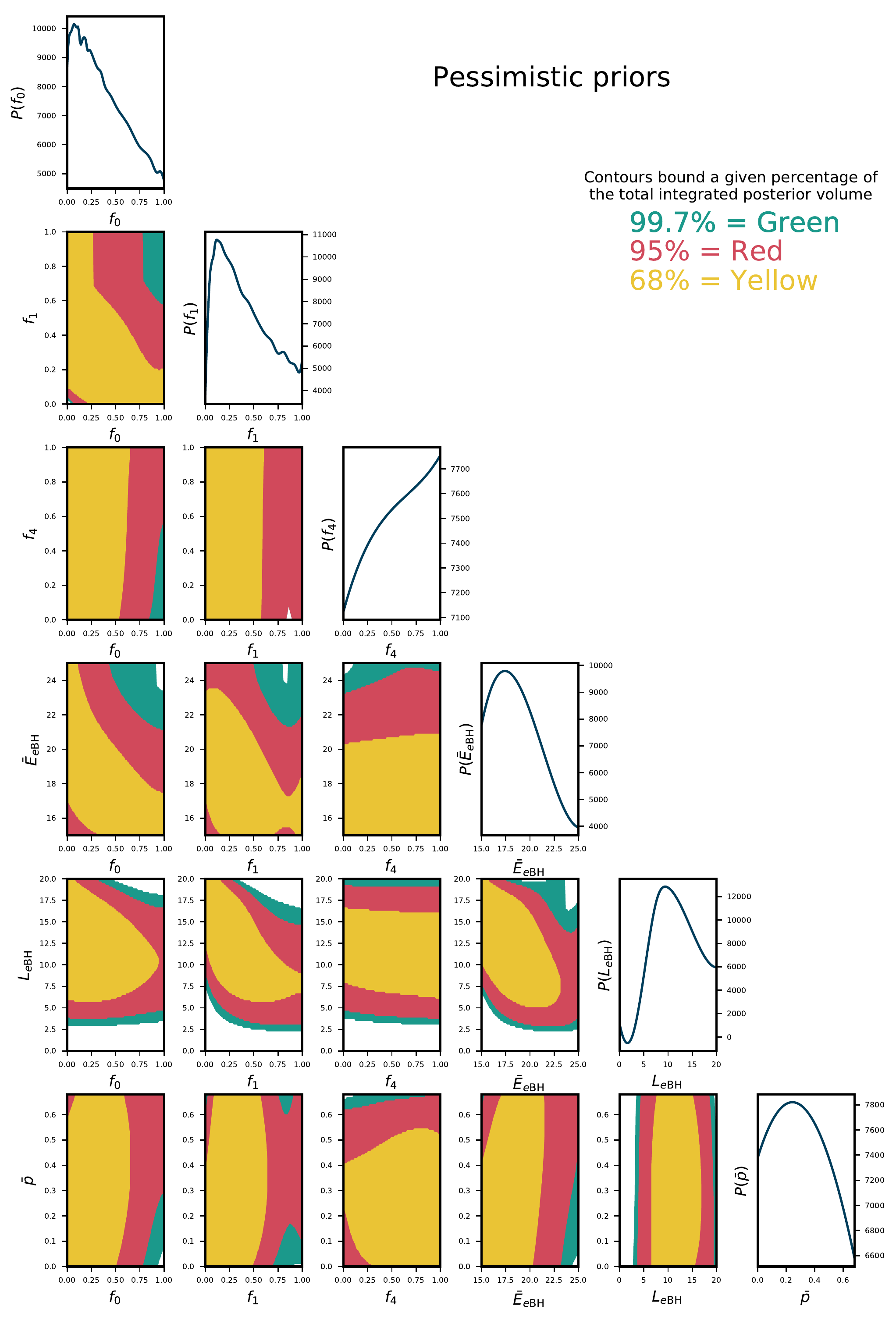}
\caption{Corner plot of of five parameters from our MCMC projection from running an experiment similar to Hyper-Kamiokande for 10 years with pessimistic priors. The contours bound a given fraction of the total integrated posterior.
For the one-dimensional histograms, $P(x)$ is the one-dimensional posterior for the parameter $x$.}
\label{fig:corner_pess}
\end{figure}

The results obtained in this paper rely on a Bayesian statistical technique called Markov Chain Monte Carlo (MCMC), which we have implemented using the $\textsc{PyMC}$ module in $\textsc{Python}$. The purpose of this MCMC algorithm is to 
obtain the ranges of values of the relevant parameters, in our case those listed in table~\ref{table_params}, for which the experimental data and theoretical prediction of signal and background are in  agreement to a certain level
of confidence. In addition, the MCMC allows any degeneracies between different theoretical parameters to be studied. 

The MCMC starts by randomly sampling each parameter according to prior distributions, which in our case are those 
listed in table~\ref{priors}. These distributions reflect the allowed and favoured values of the parameters before accounting for the experimental data, which in this case is the projected DSNB measurement by Hyper-Kamiokande.
The code then evaluates the likelihood function comparing the expected signal and background, given these sampled parameter values, with the experimental data and repeats this process many more times. 
After many such samples the algorithm attempts to sample values of the parameters which make the likelihood as large as possible, and continually attempts to maximise the likelihood as the sampling process repeats. 

In our case the MCMC samples the parameter values $200000$ times, after which the distributions of the sampled parameters no longer resemble the prior distributions, but instead reflect the posterior distribution of parameter values. 
The initial $50000$ samples are discarded as ``burn-in'', since these reflect the initial stage where the algorithm was scoping out the parameter space for the best-fit region.
In contrast with the prior
distributions, the posterior gives us the preferred values of the parameters accounting for the experimental data. The larger the value of the posterior distribution for a given set of parameter values, the better the fit of the theoretical
prediction to data.
Taking a histogram of this posterior sample from the MCMC for a given parameter and integrating gives us the confidence intervals
for the values of this parameter which most resemble the data. Likewise a two-dimensional histogram tells us the preferred regions for two parameters to given levels of confidence. 

This is what we show in figures~\ref{fig:corner_op} and \ref{fig:corner_pess}. For example, in the top three panels of figure~\ref{fig:corner_op} we have a one-dimensional histogram of all the sampled points (minus the burn-in) as a function of $f_0$, a one-dimensional
histogram as a function of $f_1$ and a two-dimensional histogram as a function of both $f_0$ and $f_1$. The one-dimensional histograms show that the best-fit values of $f_0$ and $f_1$ are both around $0.2$, but it is the two-dimensional
histogram which holds the main information. Here the shaded regions indicate contours of constant posterior value, for which the integral of the posterior distribution inside the contour is at a given fraction of the total posterior sum over all parameter values e.g. for the yellow shaded region $68\%$ of the posterior is contained within 
this contour, and so there is a $68\%$ probability to find the parameter values within this region.

\bibliographystyle{JHEP}

\providecommand{\href}[2]{#2}\begingroup\raggedright\endgroup

\end{document}